\documentclass[]{raa}
\usepackage{graphicx,times,amsmath,amssymb,natbib}             
\usepackage{natbib}
\bibpunct{(}{)}{;}{a}{}{,}

\usepackage[a4paper=true,dvipdfm=true,pagebackref=true]{hyperref}
\hypersetup{pdftitle = The title of my PDF, pdfauthor = My name, pdfsubject= The subject, pdfkeywords = keyword1 keyword2 keyword3} 
\hypersetup{colorlinks = true, linkcolor = green, anchorcolor = red, citecolor = blue, filecolor = red, pagecolor = red, urlcolor = red}
\begin{document}

\title{Broadband Spectral Fitting of Blazars using XSPEC}
   \volnopage{Vol.0 (200x) No.0, 000--000}      
   \setcounter{page}{1}          
   \author{S. Sahayanathan\inst{1} \and A. Sinha\inst{2} \and R. Misra\inst{2}}
   \institute{Astrophysical Sciences Division, Bhabha Atomic Research Centre, Mumbai - 400085, India; {\it sunder@barc.gov.in}\\
   \and
	Inter-University Center for Astronomy and Astrophysics, Post Bag 4, Pune, India\\
}

\abstract{The broadband spectral energy distribution(SED) of blazars is generally interpreted as
radiation arising from synchrotron and inverse Compton mechanisms. Traditionally, 
the underlying source parameters responsible for these emission processes, like particle energy density, magnetic field, etc.,
 are 
obtained through simple visual reproduction of the observed fluxes. However, this procedure is incapable
of providing the confidence range on the estimated parameters. In this work, we propose 
an efficient algorithm to perform a statistical fit of the observed broadband spectrum of blazars using different emission models.
Moreover, in this work we use the the observable quantities as the fit parameters, rather than
the direct source parameters which govern the resultant SED.
This significantly improves the convergence time and
eliminates the uncertainty regarding the initial guess parameters.
This approach also has an added advantage of identifying the degenerate parameters, which can be removed
by including more observable information and/or additional constraints.
A computer code developed based on this algorithm is implemented as an user-defined 
routine in the standard X-ray spectral fitting package, XSPEC. Further, we demonstrate the efficacy of the 
algorithm by fitting the well sampled SED of the blazar, 3C~279, during its gamma ray flare in 2014.
}
\authorrunning{S. Sahayanathan, A. Sinha \& R. Misra}
    \titlerunning{Broadband Spectral Fitting of Blazars}
   \maketitle
\section{Introduction}\label{s:intro}
A presence of powerful jets is one of the striking features of active galactic nuclei (AGN), with
blazars belonging to a special class where the jet is aligned close to the line of sight 
\citep{1993ARA&A..31..473A,1995PASP..107..803U}. 
The emission from blazars is predominantly non-thermal in nature and extends from
radio to gamma ray energies \citep{1996ApJ...463..444S}. Transparency 
to high energy gamma rays and a rapidly varying flux implies the jet is 
relativistic \citep{1995MNRAS.273..583D} and hence, its emission is significantly 
boosted due to
relativistic Doppler effects. Besides this non-thermal jet emission, blazar 
spectral energy distribution (SED) is often observed to have broad emission/absorption 
lines and thermal features \citep{1991ApJ...373..465F,2006ApJ...653.1089L,2011ApJ...732..116M}. 
Consisently, blazars are further subdivided into two classes, namely,
flat spectrum radio quasars (FSRQs) with broad line features and BL Lacs with
weak or no emission/absorption lines \citep{2007ApJ...662..182P}.

The broadband SEDs of blazars are characterized by a typical double hump feature which is
attributed to radiative losses encountered by a non-thermal electron distribution 
\citep{2010ApJ...716...30A}.  
The low energy component is well understood as synchrotron emission from a 
relativistic population of electrons in the jet losing its energy in a magnetic field; whereas,
the high energy emission is generally attributed to inverse Compton scattering
of soft target photons by the same electron distribution. The soft target
photons can be synchrotron photons themselves, commonly referred as synchrotron
self Compton (SSC) \citep{1981ApJ...243..700K,1985ApJ...298..114M,1989ApJ...340..181G}
and/or the other photon field from the jet environment, commonly
referred as external Compton (EC) \citep{1987ApJ...322..650B,1989ApJ...340..162M,1992A&A...256L..27D}. The most prominent external photon fields which are 
scattered off by the jet electrons via inverse Compton process are the emission
from the accretion disk (EC/disk) \citep{1993ApJ...416..458D,1997A&A...324..395B},
the reprocessed broad emission lines from broad line emitting regions (EC/BLR) \citep{1994ApJ...421..153S,1996MNRAS.280...67G} and the thermal infrared
radiation from the dusty torus (EC/IR), proposed by the unification theory \citep{1994ApJ...421..153S,2000ApJ...545..107B,2009MNRAS.397..985G}. 
The relative 
contributions of these emission processes are usually obtained by simple visual reproduction of the
broadband SED using various emissivity functions \citep{2015ApJ...803...15P,2012MNRAS.419.1660S,2013MNRAS.433.2380K,2009MNRAS.397..985G}. 
However, a proper statistical treatment of the broadband SED considering these emission processes 
has not been pursued in detail, except for a few recent works 
\citep[e.g.][]{2011ApJ...733...14M,2012ApJ...752..157Z,2014ApJS..215....5K} . 
Such a statistical treatment, besides
providing the range of source parameters which is consistent with the observation, will also
benefit us in understanding the jet environment and the possible location of the emission 
region \citep{2012ApJ...752..157Z,2013ApJ...767....8Z,2014ApJ...788..104Z,2015ApJ...807...51Z}.

The present epoch is particularly rewarding for observational astronomy due to some remarkable 
technological advancements in recent years. This has resulted in high sensitivity
experiments operating at various energy bands like, optical (e.g \emph{Hubble Space Telescope}), X-ray(e.g \emph{Swift}, 
\emph{NuSTAR}, \emph{AstroSat}) and gamma rays (e.g \emph{Fermi}, \emph{MAGIC}, \emph{VERITAS}, \emph{HESS}). With the 
availability of high quality data from these experiments through coordinated multi 
wavelength observations, we now have rich spectral information of blazars during flare as well 
as quiescent flux states \citep{2015MNRAS.450.2677C,2015A&A...573A..50A,2011ApJ...736..131A,2016A&A...591A..83S}. This development, in turn, demands more sophisticated spectral 
fitting numerical codes, involving various physical emission  models rather than simple mathematical 
functions representing a narrow range of energies \citep{2015A&A...580A.100S,2014MNRAS.444.3647B,2013A&A...557A..71R}, 
which are capable of extracting the source parameters of blazars with significant 
confidence levels. Successful reproduction of blazar SED during quiescent and different 
flaring states using such spectral fitting algorithms will help us in understanding the physics
behind blazar flares and its dynamics \citep{2015ApJ...803...15P,2014MNRAS.442..131K,2009MNRAS.397..985G}.

The main challenge encountered while developing the algorithms for broadband spectral fitting of
blazars, involving different physical emission models, is the numerical intensiveness. The presence
of multiple integrations in different emissivity formulae require a large number of nested loops making the 
algorithms computationally intensive \citep{2011ApJ...733...14M,2012ApJ...752..157Z,2014ApJS..215....5K}. 
In addition, a complex dependence of the source parameters
on the observed flux levels makes the algorithms wander considerably in the 
parameter space, eventually slowing down the fitting process \citep{1986rpa..book.....R,1970RvMP...42..237B,1993ApJ...416..458D}. 
However, thanks to the availability of modern high speed computers with multi core processors and 
optimized numerical algorithms directed towards effective utilization of resources, one can now perform this
spectral fitting procedure relatively faster. 

The attempt to perform a statistical fitting of blazar SED was first initiated by \cite{2011ApJ...733...14M},
where the authors fitted the multi-epoch, broadband SED of Mrk 421 using synchrotron 
and SSC processes. The fitting was performed using $\chi^2$ minimization technique incorporating 
Levenberg-Marquardt algorithm \citep{1992nrfa.book.....P}. For such algorithms, convergence to the actual minima is strongly 
dependent on the initial guess values of the source parameters. However, the non-linear dependence 
of the source parameters with different emissivity functions often makes it hard/impossible to 
choose the right set of initial guess values to begin with. This may eventually lead the minimization 
algorithm to descent towards unphysical parameter space. Alternatively, a novel approach was proposed by
\cite{2012ApJ...752..157Z} where the authors used the observed information to extract most of the source
parameters \citep{1998ApJ...509..608T}. The source magnetic field and the jet Doppler factor are finally 
obtained through $\chi^2$ minimization.
This approach has significantly eased the problem of choosing the initial guess values. 
Recently, \cite{2014ApJS..215....5K} added EC/IR and EC/BLR processes along 
with synchrotron and SSC processes and performed a spectral fitting for the SED of 28 low energy peaked BL Lac 
objects. For each source, they generated the SED corresponding to broad range of parameters and
calcuated the $\chi^2$. The best fit parameters and their errors were estimated from this $\chi^2$
space. However, such algorithms are inefficient and excessive computational time 
forced the authors to freeze certain parameters.

In this work, we develop an algorithm considering synchrotron, SSC and EC mechanisms, 
	to fit the broadband SED of blazars using the 
	standard X-ray spectral fitting package \emph{XSPEC} \citep{1996ASPC..101...17A}). XSPEC is primarily developed 
to obtain the X-ray fluxes from the source by convolving a source spectral model function 
with the detector response matrix of the satellite based X-ray telescopes. It employs the 
Levenberg-Marquardt algorithm to fit the observed photon counts with the model spectrum
and produce the ``most probable'' flux of the source. The software package also provides 
the flexibility to add 
user defined spectral models (local models) and fit with the observed photon counts. 
We developed separate additive local models  
for synchrotron, SSC and EC processes which can be added according to the necessity. 
Rather than fitting the direct source parameters governing the underlying spectrum, we 
fit the observed spectrum. This ensures faster convergence and removes the problem of guessing initial values.
The numerical codes for various
emissivities are significantly optimized to reduce the machine run time. An added advantage of
using XSPEC spectral fitting package, besides being well optimized and widely tested,
is that it allows us to fit the photon counts within the energy bins rather than the fluxes at their mean
energy. The XSPEC routines are finally applied on the well studied FSRQ, 3C 279, as a test
case. The choice of 3C 279 is mainly driven by the fact that the non-thermal emission dominates 
its entire SED, availability of sufficient multi wavelength data and the need for EC process to reproduce
its gamma ray observation \citep{2012MNRAS.419.1660S}. 

The paper is organised as follows: in \S  \ref{sec:emission_models}, 
we describe the different emission models relevant for the broadband 
spectral fitting of the non thermal emission from blazars. Here, we derive the emissivity
formulae for the synchrotron and inverse Compton processes and show their relation with the 
observed spectral information. In \S  \ref{sec:xspecfit}, we present the proposed spectral
fitting procedure using XSPEC and its application on 3C\, 279, and in \S  \ref{sec:discussion},
we discuss the implications and advantages of the developed spectral fitting algorithm.
A cosmology with $\Omega_m = 0.3$, $\Omega_\Lambda = 0.7$ and $H_0 = 70\,\textrm{km}\,\textrm{s}^{-1}\,\textrm{Mpc}^{-1}$ is used in
this work.

\section[]{Blazar Jet Emission Models}\label{sec:emission_models}
We model the non-thermal emission from the blazar jet to originate from a spherical region of radius $R$,
moving down the jet with bulk Lorentz factor $\Gamma$ at an angle $\theta$ with respect to the line of sight
of the observer.
The emission region is filled with a broken power law electron 
distribution, given by
\begin{align} \label{eq:broken}
	N(\gamma)\,d\gamma = \left\{
\begin{array}{ll}
	K\,\gamma^{-p}\,d\gamma&\textrm{for}\quad \mbox {~$\gamma_{\rm min}<\gamma<\gamma_b$~} \\
	K\,\gamma_b^{q-p}\gamma^{-q}\,d\gamma&\textrm{for}\quad \mbox {~$\gamma_b<\gamma<\gamma_{\rm max}$~}
\end{array}\quad {\rm cm}^{-3}
\right.
\end{align}
undergoing synchrotron loss due to a tangled magnetic field, $B$, and inverse Compton losses by
scattering off low energy photons. Here, $\gamma$ ($=\frac{E}{m_ec^2}$) is the dimensionless energy 
with $m_e$ the mass of electron and $c$ being the velocity of light, $K$ the normalisation factor, 
$\gamma_b$ is the break energy and $p$ and $q$ are the low and high energy electron spectral indices. 
The target photons for the inverse Compton scattering are 
synchrotron photons (SSC) and an isotropic blackbody photon field 
at temperature $T_*$ external to jet \footnote{Quantities with subscript $*$ are measured in the 
co-moving frame where the parent galaxy is at rest;
whereas, the rest of the quantities are measured in emission region frame where the electron distribution is homogeneous, unless mentioned 
otherwise.}. 

\subsection{Synchrotron Specific Intensity}
The synchrotron emissivity due to an isotropic electron distribution losing its energy
in a tangled magnetic field, $B$, is given by \citep{1986rpa..book.....R}
\begin{align}\label{eq:syn_emiss}
	j_{\rm syn}(\nu)= \frac{1}{4\pi}\int\limits_{\gamma_{\rm min}}^{\gamma_{\rm max}} P_{\rm syn}(\gamma,\nu)\,N(\gamma)\,d\gamma
	\quad {\rm erg/cm}^3{\rm /s/Hz/Sr}
\end{align}
where, $P_{\rm syn}(\gamma,\nu)$ is the pitch angle averaged single particle emissivity, given by
\begin{align}
	P_{\rm syn}(\gamma,\nu)=\frac{\sqrt{3}\pi e^3 B}{4m_ec^2} F\left(\frac{\nu}{\nu_c}\right)
	\quad {\rm erg/s/Hz}
\end{align}
with, 
\begin{align}
\nu_c=\frac{3\gamma^2 e B}{16 m_e c} \quad Hz
\end{align}
and synchrotron power function \citep{1980panp.book.....M}
\begin{align}
	F(x) &= x\int\limits_x^\infty K_{5/3}(\xi)\, d\xi \nonumber \\
	&\approx 1.8 \, x^{1/3}\, e^{-x}
\end{align}
Here, $K_{5/3}$ is the modified Bessel function of order $5/3$. The function $F(x)$ 
has a single peak at $x\approx0.29$. 
At the optically thick regime, synchrotron photons are self absorbed and the absorption
coefficient is given by \citep{1991MNRAS.252..313G,1999MNRAS.306..551C}
\begin{align} \label{eq:ssa}
\kappa(\nu) = -\frac{1}{8 \pi m_e \nu^2}\int\limits_{\gamma_{\rm min}}^{\gamma_{\rm max}}
\frac{N(\gamma)}{\gamma \sqrt{\gamma^2-1}} \frac{d}{d\gamma}\left[\gamma\sqrt{\gamma^2-1}\,P_{\rm syn}(\gamma,\nu)\right] \quad cm^{-1}
\end{align}
Using the emissivity and absorption coefficient, equation (\ref{eq:syn_emiss}) and (\ref{eq:ssa}), the synchrotron 
specific intensity can be obtained from the radiative transfer equation as \citep{1986rpa..book.....R}
\begin{align}
	I_{\rm syn}(\nu)=\frac{j_{\rm syn}(\nu)}{\kappa(\nu)}\left[1-e^{-\kappa(\nu)R}\right]
	\quad {\rm erg/cm}^2{\rm /s/Hz/Sr}
\end{align}
For optically thin regime, $I_{\rm syn}(\nu)\approx j_{\rm syn}(\nu)\, R$.

Alternatively, an approximate analytical solution of the synchrotron emissivity can be obtained
by assuming the single particle emissivity as \citep{1991pav..book.....S}
\begin{align}\label{eq:syn_power_approx}
P_{\rm syn}(\gamma,\nu) = \frac{4}{3}\beta^2\gamma^2c\sigma_TU_B\Phi_\nu(\gamma) 
\end{align}
where, $\beta$ ($=\frac{v}{c}$) is the dimensionless velocity of the emitting electron, $\sigma_T$ is the
Thomson cross section and the spectral function $\Phi_\nu(\gamma)$ satisfies the relation
\begin{align}
\int\limits_0^\infty \Phi_\nu(\gamma) \, d\nu = 1
\end{align}
In case of synchrotron emissivity due to non-thermal distribution of electrons, the narrow shape of $F(x)$ let us
approximate the function $\Phi_\nu(\gamma)$ as a $\delta$-function
\begin{align}
\Phi_\nu(\gamma) \to \delta(\nu-\gamma^2\nu_L)
\end{align}
where, the Larmor frequency $\nu_L = \frac{eB}{2\pi m_ec}$. Using this approximation 
on equation (\ref{eq:syn_power_approx}) and the $\delta$-function property 
\begin{align}\label{eq:delfn}
	\delta[f(x)]=\sum_i \frac{\delta(x-x_i)}{\left| \frac{df}{dx}\right|_{x=x_i}}
\end{align}
with $x_i$'s being the roots of $f(x)$, the synchrotron emissivity can be obtained as  
as\footnote{$\sim$ hat represents approximate analytical estimates} \citep{2012MNRAS.419.1660S}
\begin{align}\label{eq:syn_approx}
	\tilde{j}_{\rm syn}(\nu)\approx\frac{\sigma_TcB^2}{48 \pi^2}\nu_L^{-\frac{3}{2}}
  N\left(\sqrt{\frac{\nu}{\nu_L}}\right)\nu^{\frac{1}{2}}
  \quad {\rm erg/cm}^3{\rm /s/Hz/Sr}
\end{align}

\subsection{SSC Emissivity}

The polarization angle averaged differential Compton cross section, in the rest frame 
of the scattering electron\footnote{Quantities with prime are measured in the
electron rest frame}, is given by the Klein-Nishina formula \citep{1970RvMP...42..237B}
\begin{align} \label{eq:knexact}
	\frac{d^2\sigma}{d\nu_s'\,d\Omega_s'} = \frac{r_e^2}{2}
	\left(\frac{\nu_s'}{\nu_i'}\right)^2
	\left(\frac{\nu_i'}{\nu_s'}+\frac{\nu_s'}{\nu_i'}
	-1+\textrm{cos}\,\psi'^2\right) 
	\delta\left[\nu_s'-\frac{\nu_i'}{1+\frac{h\nu_i'}{m_ec^2}(1-\textrm{cos}\,\psi')}\right]
	\quad {\rm cm}^2{\rm /Sr/Hz}
\end{align}
where, $\nu_i'$ and $\nu_s'$ are the frequency of the incident and the scattered
photon, $\psi'$ is the angle between their direction, $h$ is the Planck's constant and $r_e$ 
is the classical electron radius. For the case of elastic scattering,
$\nu_s'\approx\nu'_i$ and the equation (\ref{eq:knexact}) reduces to the classical Thomson
limit
\begin{align}
	\frac{d^2\sigma}{d\nu_s'\,d\Omega_s'} \approx \frac{r_e^2}{2}(1+\textrm{cos}\,\psi'^2)\,
\delta(\nu_s'-\nu_i')
\end{align}
The single particle Compton emissivity due to scattering of the isotropic 
synchrotron photons
can then be obtained from Klein-Nishina formula as \citep{1970RvMP...42..237B,1968PhRv..167.1159J}
\begin{align}
	P_{\rm ssc}(\gamma,\nu_s) =\frac{3\pi \sigma_T \nu_s}{\gamma^2}\int\limits_{x_1}^{x_2} 
	\frac{I_{\rm syn}(\nu_i)}{\nu_i^2} f(\nu_i,\nu_s,\gamma)\, d\nu_i
	\quad {\rm erg/s/Hz}
\end{align}
where,
\begin{align}
	x_1 &= MAX\left[\nu_{\rm syn}^{\rm min},\frac{\nu_s}
	{4\gamma^2\left(1-\frac{h\nu_s}{\gamma m_ec^2}\right)}\right]; 
	\quad \nu_{\rm syn}^{\rm min} \approx 1.29\times10^6\gamma_{\rm min}^2B \\
	x_2 &= MIN\left[\nu_{\rm syn}^{\rm max},\frac{\nu_s}
	{\left(1-\frac{h\nu_s}{\gamma m_ec^2}\right)}\right]; \quad
	\quad \nu_{\rm syn}^{\rm max} \approx 1.29\times10^6\gamma_{\rm max}^2B 
\end{align}
and
\begin{align}
	f(\nu_i,\nu_s,\gamma)=
	2q\,\textrm{log}\,q+(1+2q)(1-q)+\frac{(\zeta q)^2(1-q)}{2(1+\zeta q)}
\end{align}
with 
\begin{align}
        \zeta=\frac{4\gamma h\nu_i}{m_ec^2}\quad \textrm{and}\quad 
	q = \frac{\nu_s}{4\nu_i\gamma^2\left(1-\frac{h\nu_s}{\gamma m_e c^2}\right)}
	\nonumber
\end{align}
Finally, the SSC emissivity due to the electron distribution given in equation (\ref{eq:broken})
will be
\begin{align}\label{eq:ssc_emiss}
	j_{\rm ssc}(\nu)= \frac{1}{4\pi}\int\limits_{\gamma_{\rm min}}^{\gamma_{\rm max}} P_{\rm ssc}(\gamma,\nu)\,N(\gamma)\,d\gamma
	\quad {\rm erg/cm}^3{\rm /s/Hz/Sr}
\end{align}

Similar to the synchrotron case, an approximate analytical solution of SSC emissivity, 
happening in the Thomson regime, can be obtained by considering the single particle 
emissivity as \citep{2012MNRAS.419.1660S}
\begin{align}
	P_{\rm ssc}(\gamma,\nu) = \frac{4}{3}\beta^2\gamma^2c\sigma_T
	\int\limits_{\nu_{\rm syn}^{\rm min}}^{\nu_{\rm syn}^{\rm max}} U(\xi) \; d\xi \;
\Psi_\nu(\xi,\gamma)
\end{align}
where, 
\begin{align}
U_{\rm ph} = \int\limits_{\nu_{\rm syn}^{\rm min}}^{\nu_{\rm syn}^{\rm max}} U(\xi) d\xi
\quad {\rm erg/cm}^3
\end{align} 
is the energy density of the synchrotron photons and the function $\Psi_\nu(\xi,\gamma)$
satisfies the condition
\begin{align}
\int\limits_0^\infty \Psi_\nu(\xi,\gamma) d\nu = 1
\end{align}
Since the scattered photon frequency in the Thomson regime is $\gamma^2\xi$ approximately,
we can express $\Psi_\nu(\xi,\gamma)$ as
\begin{align}
\Psi_\nu(\xi,\gamma) \to \delta(\nu-\gamma^2\xi)
\end{align}
From equation (\ref{eq:ssc_emiss}), the SSC emissivity will then be
\begin{align}
	\tilde{j}_{\rm ssc}(\nu) \approx \frac{1}{3\pi}c\sigma_T\int\limits_{\gamma_{\rm min}}^{\gamma_{\rm max}}
U\left(\frac{\nu}{\gamma^2}\right)N(\gamma)\, d\gamma \nonumber
\end{align}
Expressing $U(\nu)= \frac{4\pi R}{c} j_{\rm syn}(\nu)$ and using equation (\ref{eq:syn_approx})
we get
\begin{align}
	\tilde{j}_{\rm ssc}(\nu)\approx
\frac{R c}{36\pi^2}\sigma_T^2B^2\nu_L^{-\frac{3}{2}}\nu^{\frac{1}{2}}
\int\limits_{\gamma_{\rm min}}^{\gamma_{\rm max}} \frac{d\gamma}{\gamma}N\left(\frac{1}{\gamma}\sqrt{\frac{\nu}{\nu_L}}\right)
N(\gamma)
\end{align}
For the case of non-thermal electron distribution, given by equation(\ref{eq:broken}), we obtain
\begin{align}\label{eq:ssc_approx}
	\tilde{j}_{\rm ssc}(\nu) \approx \frac{R c}{36\pi^2}K^2\sigma_T^2B^2\nu_L^{-\frac{3}{2}}
\nu^{\frac{1}{2}}\mathnormal{f}(\nu)
\end{align}
Here,
\begin{align}
\mathnormal{f}(\nu) = \, &\left[\left(\frac{\nu}{\nu_L}\right)^{-\frac{p}{2}}
	\textrm{log}\left(\frac{\gamma_1}{\gamma_2}\right) 
+\frac{\gamma_b^{(q-p)}}{q-p}\left(\frac{\nu}{\nu_L}\right)^{-\frac{q}{2}}
\nonumber \right.\\ & \left.\times
(\gamma_1^{(q-p)}-\gamma_{\rm min}^{(q-p)})
\Theta\left(\frac{1}{\gamma_b}\sqrt{\frac{\nu}{\nu_L}}-\gamma_{\rm min}\right)
\right]\Theta(\gamma_2-\gamma_1)
\nonumber \\ 
&+\left[\gamma_b^{2(q-p)}\left(\frac{\nu}{\nu_L}\right)^{-\frac{q}{2}}
	\textrm{log}\left(\frac{\gamma_4}{\gamma_3}\right) +
\frac{\gamma_b^{(q-p)}}{q-p}\left(\frac{\nu}{\nu_L}\right)^{-\frac{p}{2}}
\nonumber \right. \\  &\left. \times
(\gamma_4^{(p-q)}-\gamma_{\rm max}^{(p-q)})
\Theta\left(\gamma_{\rm max}-\frac{1}{\gamma_b}\sqrt{\frac{\nu}{\nu_L}}\right)\right]
\Theta(\gamma_4-\gamma_3)
\end{align}
with $\Theta$ being the Heaviside function and 
\begin{align}
	\gamma_1 &= {\rm MAX}\left(\gamma_{\rm min},\frac{1}{\gamma_b}\sqrt{\frac{\nu}{\nu_L}}\right) \nonumber \\
	\gamma_2 &= {\rm MIN}\left(\gamma_b,\frac{1}{\gamma_{\rm min}}\sqrt{\frac{\nu}{\nu_L}}\right) \nonumber \\
	\gamma_3 &= {\rm MAX}\left(\gamma_b,\frac{1}{\gamma_{\rm max}}\sqrt{\frac{\nu}{\nu_L}}\right) \nonumber \\
	\gamma_2 &= {\rm MIN}\left(\gamma_{\rm max},\frac{1}{\gamma_b}\sqrt{\frac{\nu}{\nu_L}}\right) 
\end{align}

\subsection{EC Emissivity}
The EC emissivity for the case of relativistic electrons with $\gamma\gg1$  
can be estimated following the procedure described in \cite{1993ApJ...416..458D} and \cite{2009herb.book.....D}. 
Under this case, the direction of the scattered photon ($\Omega_s$), in the frame of the emission region, 
can be approximated to be that of the electron itself and the differential 
Compton cross section in the emission region frame can be written as
\begin{align}\label{eq:crosserf}
	\frac{d^2\sigma}{d\nu_s\,d\Omega_s} = \delta(\Omega_s-\Omega_e)
	\oint d\Omega_s' \left(\frac{d\nu_s'}{d\nu_s}\right)\,
	\frac{d^2\sigma}{d\nu_s'\,d\Omega_s'}
\end{align}
where, $\Omega_e$ is the direction of the scattering electron and $\nu_s$ is the frequency of the 
scattered photon. Again, $\gamma\gg1$
also allows one to approximate the direction of the incident photon in the electron 
rest frame to be opposite
to the direction of the electron (head-on approximation). Hence, the cosine of the angle
between the incident and the scattered photon,
cos$\,\psi'\approx -\mu_s'$ where, $\mu_s'$ is the cosine of the angle between 
the direction of electron and the scattered photon. The quantities $\mu_s'$ and  
$\nu_s'$ are related to the corresponding quantities in the frame of emission
region as
\begin{align}
	\label{eq:musp}
	\mu_s'&=\frac{\mu_s-\beta}{1-\beta\mu_s} \\
	\label{eq:nusp}
	\nu_s'&=\nu_s\,\gamma(1-\beta\mu_s)
\end{align}
From equations (\ref{eq:musp}) and (\ref{eq:nusp}) we get
\begin{align}\label{eq:omegarat}
	\frac{d\Omega_s'}{d\Omega_s} = \left(\frac{\nu_s}{\nu_s'}\right)^2
\end{align}
Using equations (\ref{eq:nusp}) and (\ref{eq:omegarat}), (\ref{eq:crosserf}) 
can be expressed as
\begin{align} \label{eq:crosser}
	\frac{d^2\sigma}{d\nu_s\,d\Omega_s} = \delta(\Omega_s-\Omega_e)
	\oint d\Omega_s \left(\frac{\nu_s}{\nu_s'}\right)\,
	\frac{d^2\sigma}{d\nu_s'\,d\Omega_s'}
\end{align}
The above equation relates the differential Compton cross section between the 
emission region and the electron rest frame. 
The $\delta$-function in equation (\ref{eq:knexact}) can be modified using equation (\ref{eq:delfn})
as
\begin{align}\label{eq:deltachange}
\delta\left[\nu_s'-\frac{\nu_i'}{1+\frac{h\nu_i'}{m_ec^2}(1+\mu_s')}\right]=
\frac{1}{\nu_s\left|\frac{h\nu_s}{m_ec^2}-\gamma\beta\right|}
\,\delta\left[\mu_s-\frac{1+\frac{h\nu_i'}{m_ec^2}(1-\beta)-\frac{\nu_i'}{\gamma\nu_s}}
{\beta-\frac{h\nu_i'}{m_ec^2}(1-\beta)}\right]
\end{align}
and the differential Compton cross section in the emission region frame, equation (\ref{eq:crosser}), 
can be expressed as
\begin{align}\label{eq:crosser1}
	\frac{d^2\sigma}{d\nu_s\,d\Omega_s} = 
	\frac{\pi r_e^2}{\gamma \nu_i'}
	\,\delta(\Omega_s-\Omega_e) \,\Xi(\gamma,\nu_s,\nu_i');\quad
	\frac{\nu_i'}{2\gamma}\le\nu_s\le\frac{2\gamma\nu_i'}{1+2\frac{h\nu_i'}{m_ec^2}}
\end{align}
where,
\begin{align}
	\Xi(\gamma,\nu_s,\nu_i')&=\left[y+\frac{1}{y}+\frac{\nu_s^2}{\gamma^2\nu_i'^2y^2}-
	\frac{2\nu_s}{\gamma\nu_i'y}\right] \quad \textrm{and} \quad
	y=1-\frac{h\nu_s}{\gamma m_ec^2}
\end{align}
The knowledge of differential Compton cross section lets us the write the inverse 
Compton emissivity as
\begin{align}\label{eq:emissic}
	j_{\rm ic}(\nu,\Omega) = c\,\nu \int\limits_0^\infty d\nu_i\oint d\Omega_i
	\int\limits_1^\infty d\gamma \oint d\Omega_e \,(1-\beta\,\mu_{ie})
	\,N_e(\gamma,\Omega_e)\, \frac{U_{\rm ph}(\nu_i,\Omega_i)}{\nu_i}\,\frac{d^2\sigma}{d\nu\,d\Omega}
	\quad {\rm erg/cm}^3{\rm /s/Hz/Sr}
\end{align}
where, $\Omega_i$ is the direction of the incident photon, $N_e(\gamma,\Omega_e)$ is the 
scattering electron number density (cm$^{-3}$Sr$^{-1}$), $U_{\rm ph}(\nu_i,\Omega_i)$ is the 
target photon energy density (erg cm$^{-3}$ Sr$^{-1}$) and $\mu_{ie}$ is the cosine of the angle between the 
incident photon and the scattering electron, given by
\begin{align}
	\mu_{ie}=\mu_i\mu_e+\sqrt{(1-\mu_i^2)(1-\mu_e^2)} \,\textrm{cos}(\phi_i-\phi_e)
\end{align}
with $\mu_i$ and $\mu_e$ being the cosine of the angles subtended by the incident photon 
and the scattering electron with the jet axis and $\phi_i$ and $\phi_e$ are the corresponding 
azimuthal angles. Substituting equation (\ref{eq:crosser1}) on (\ref{eq:emissic}), we
get
\begin{align}
	j_{\rm ic}(\nu,\Omega)&= \frac{3}{8} \nu c \sigma_T
	\int\limits_0^\infty d\nu_i\oint d\Omega_i
	\int\limits_1^\infty d\gamma \,(1-\beta\,\mu_{ie})
	\,\frac{N_e(\gamma,\Omega)}{\gamma}\, 
	\frac{U_{\rm ph}(\nu_i,\Omega_i)}{\nu_i'\nu_i}
	\,\Xi(\gamma,\nu,\nu_i') \nonumber \\
&= \frac{3}{8} \nu c \sigma_T
	\int\limits_0^\infty d\nu_i\oint d\Omega_i
	\int\limits_1^\infty d\gamma 
	\,\frac{N_e(\gamma,\Omega)}{\gamma^2}\, 
	\frac{U_{\rm ph}(\nu_i,\Omega_i)}{\nu_i^2}
	\,\Xi(\gamma,\nu,\nu_i')
\end{align}
where, we have used $\nu_i'=\nu_i \,\gamma(1-\beta\,\mu_{ie})$.

In case of EC process, the energy density of the target photon in the AGN frame can be 
transformed to the frame of emission region using Lorentz invariance \citep{1986rpa..book.....R}
\begin{align}
	\frac{U_{\rm ph}(\nu_i,\Omega_i)}{\nu_i^3} = \frac{U_{\rm ph*}(\nu_{i*},\Omega_{i*})}{\nu_{i*}^3}
\end{align}
where, $\nu_{i*}$ ($=\nu_i\Gamma(1+\,\beta_{\Gamma}\,\mu_i)$) is the frequency of the 
photon in the 
AGN frame and $\beta_{\Gamma}$ ($=\sqrt{1-1/\Gamma^2}$) is the dimensionless bulk 
velocity of the emission region down the jet. Hence, for the case of an isotropic external photon field,
the EC emissivity will be 
\begin{align}\label{eq:ecemiss1}
	j_{\rm ec}(\nu,\Omega)= \frac{3}{32\pi} \frac{\nu c \sigma_T}{\Gamma^2}
\int\limits_0^\infty d\nu_{i*}
	\int\limits_1^\infty d\gamma 
	\,\frac{N_e(\gamma,\Omega)}{\gamma^2}\, 
	\frac{U_{\rm ph*}(\nu_{i*})}{\nu_{i*}^2}
	\oint d\Omega_i \frac{1}{(1+\,\beta_{\Gamma}\,\mu_i)^2}
	\,\Xi(\gamma,\nu,\nu_i')
\end{align}
For $\Gamma\gg1$, relativistic beaming will cause the external photon to arrive in a direction 
opposite to the jet flow within a narrow cone of semi vertical angle $1/\Gamma$. Hence,
$\mu_{ie} \approx-\,\mu$ and being independent of $\mu_i$, $\Xi$ can be excluded from the
last solid angle integration. 
Here, $\mu$ is the cosine of the angle between the scattered photon and the 
jet direction. The integration over solid angle can then be performed analytically
\begin{align}
	\oint \frac{d\Omega_i}{(1+\,\beta_{\Gamma}\,\mu_i)^2} = 4\pi\beta_{\Gamma}\Gamma^2
\end{align}
and equation (\ref{eq:ecemiss1}) will be reduced to
\begin{align}\label{eq:ecemiss2}
	j_{\rm ec}(\nu,\Omega)= \frac{3}{8} \nu c \beta_{\Gamma} \sigma_T
\int\limits_0^\infty d\nu_{i*}
	\int\limits_1^\infty d\gamma 
	\,\frac{N_e(\gamma,\Omega)}{\gamma^2}\, 
	\frac{U_{\rm ph*}(\nu_{i*})}{\nu_{i*}^2}
	\,\Xi(\gamma,\nu,\nu_i')
\end{align}
For isotropic broken power-law distribution of electrons given in equation (\ref{eq:broken}),
we get
\begin{align}\label{eq:ecemiss3}
	j_{\rm ec}(\nu,\Omega)= \frac{3}{32\pi} \nu c \beta_{\Gamma} \sigma_T
\int\limits_0^\infty d\nu_{i*}
\int\limits_{\gamma_{\rm min}}^{\gamma_{\rm max}} d\gamma 
	\,\frac{N(\gamma)}{\gamma^2}\, 
	\frac{U_{\rm ph*}(\nu_{i*})}{\nu_{i*}^2}
	\,\Xi(\gamma,\nu,\nu_i')
\end{align}
and 
\begin{align}
	\nu_i'&\approx \Gamma\gamma(1+\beta\,\mu) \nu_{i*}
\end{align}
where, we have assumed $\nu_i\approx\Gamma\nu_{i*}$ (head-on). 
Since $\mu$ corresponds to the viewing angle $\theta$ in the AGN frame, we can 
express the former in terms of the latter as
\begin{align}\label{eq:mu_er}
	\mu&=\frac{\textrm{cos}\,\theta - \beta_{\Gamma}}{1-\beta_{\Gamma}\textrm{cos}\,\theta} \nonumber \\
	&=\delta_D \Gamma(\textrm{cos}\,\theta - \beta_{\Gamma})
\end{align}
where, $\delta_D$ ($=[\Gamma(1-\beta_{\Gamma}\textrm{cos}\,\theta)]^{-1}$) is the Doppler factor.

For the case of monochromatic external photon field, an approximate analytical solution for
EC emissivity can be obtained when the scattering process is in Thomson regime \citep{1995ApJ...446L..63D}.
Transformation of the scattered photon frequency from electron rest frame to emission region
frame will give us $\nu_s=\nu_i\,\gamma^2(1-\beta\textrm{cos}\,\psi)$ \citep{1986rpa..book.....R} and under head-on
approximation, the differential Compton cross section in the frame of emission region can be 
written as
\begin{align}
	\frac{d^2\sigma}{d\nu_s\,d\Omega_s} \approx 
	\sigma_T \,\delta(\Omega_s-\Omega_e) \,\delta[\nu_s-\nu_i\,\gamma^2(1-\beta\mu_{ie})]
\end{align}
For $\Gamma\gg1$ and $\gamma\gg1$, the incident photons travel opposite to the jet axis and we can approximate
$1-\beta\mu_{ie}\to1+\mu_e$ and $U_{\rm ph}(\nu_i,\Omega_i)\approx U_{\rm ph}(\nu_i)\,\delta(\Omega_i)$.
Hence, the inverse Compton emissivity equation (\ref{eq:emissic}) will be
\begin{align}
	\tilde{j}_{\rm ec}(\nu,\Omega) &\approx c\,\nu \sigma_T \int\limits_0^\infty d\nu_i
	\int\limits_1^\infty d\gamma \,(1+\mu)
	\,N_e(\gamma,\Omega)\, \frac{U_{\rm ph}(\nu_i)}{\nu_i}\, 
	\,\delta[\nu-\nu_i\,\gamma^2(1+\mu)]\\
	&=\frac{1}{2}c\ \sigma_T \sqrt{\nu}\int\limits_0^\infty d\nu_i
	\nu_i^{-3/2}\sqrt{1+\mu} 
	\,N_e\left[\sqrt{\frac{\nu}{\nu_i(1+\mu)}},\Omega\right]\, U_{\rm ph}(\nu_i)
\end{align}
where, we have used the $\delta$-function property equation (\ref{eq:delfn}), to perform the
integration over $\gamma$. Since $U_{\rm ph}(\nu_i)\,d\nu_i=\Gamma^2U_{\rm ph*}(\nu_{i*})\,d\nu_{i*}$ and
$\nu_i=\Gamma\nu_{i*}$, we get
\begin{align}
	\tilde{j}_{\rm ec}(\nu,\Omega)=\frac{1}{2}c\ \sigma_T \sqrt{\Gamma\nu(1+\mu)}
	\int\limits_0^\infty d\nu_{i*}
	\nu_{i*}^{-3/2} 
	\,N_e\left[\sqrt{\frac{\nu}{\Gamma\nu_{i*}(1+\mu)}},\Omega\right]\, U_{\rm ph*}(\nu_{i*})
\end{align}
For a monochromatic external photon field, $U_{\rm ph*}(\nu_{i*})=U_*\delta(\nu_{i*}-\overline{\nu}_*)$ at
frequency $\overline{\nu}_*$, and for 
an isotropic electron distribution we get
\begin{align}\label{eq:ec_approx}
	\tilde{j}_{\rm ec}(\nu,\Omega)=\frac{c\sigma_T U_*}{8\pi\overline{\nu}_*} 
	\sqrt{\frac{\Gamma\nu(1+\mu)}{\overline{\nu}_*}}
	\,N_e\left[\sqrt{\frac{\nu}{\Gamma\overline{\nu}_{*}(1+\mu)}}\right]
\end{align}
and from equation(\ref{eq:mu_er}),
\begin{align}
	\Gamma(1+\mu)=\delta_D\left(\frac{\textrm{cos}\,\theta+1}{1+\beta_{\Gamma}}\right)
\end{align}
It should be noted here that an external photon field of blackbody type can be 
approximated as a monochromatic, owing to broad spectral range of EC emissivity resulting
from a power law electron distribution.

\subsection{Observed Flux}\label{sec:obs_approx}
\begin{figure}
\centerline{\includegraphics[angle=-90,width=12cm]{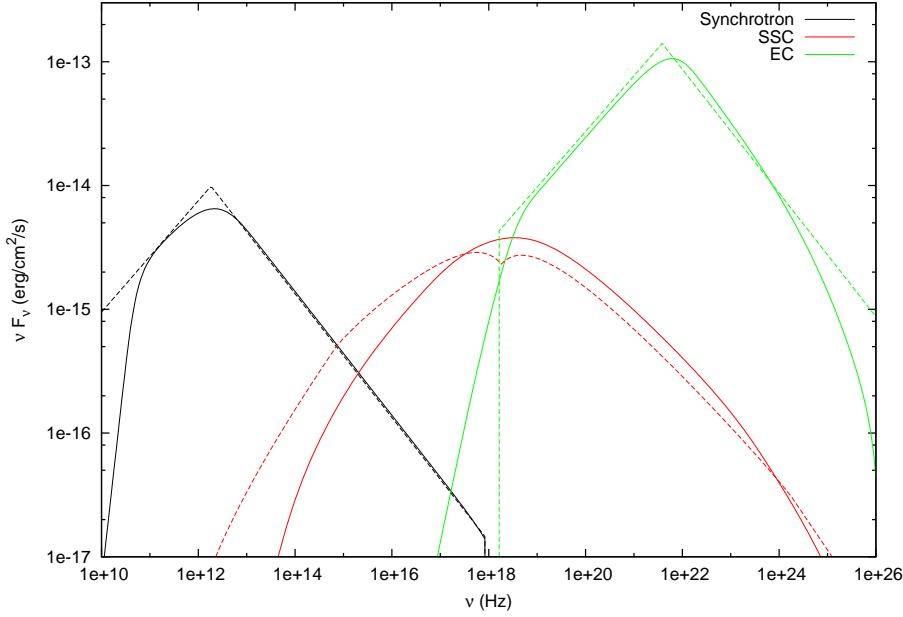}}
\caption{The derived synchrotron, SSC and EC model spectrum (solid lines) with their approximate 
analytical equivalents (dashed lines). The model SED corresponds to the following source 
parameters: $z=0.536$, $p=0.55$, $q=1.5$, $K=1\times10^5$, $\gamma_{\rm min}=10$, 
$\gamma_{\rm max}=5\times10^5$, $\gamma_b=10^3$, $B=0.1\, G$, $\Gamma=10$, $\delta_D=10$, 
$\overline{\nu}_*=5.86\times10^{13}$ Hz equivalent to temperature $1000$ K,   
$U_*=7.57\times10^{-5}$ erg/cm$^3$ and $R=10^{16}$ cm
\label{fig:approx}}
\end{figure}

The flux received by the observer due to synchrotron and inverse Compton 
emission processes can be obtained
from their corresponding emissivities. After accounting for the relativistic Doppler boosting
and cosmological effects, the observed flux\footnote{Quantities with subscript 'obs' are measured in the
observer's frame} at frequency $\nu_{\rm obs}$ in the direction
$\Omega_{\rm obs}$ will be \citep{1984RvMP...56..255B,1995ApJ...446L..63D}
\begin{align}\label{eq:obs_flux}
	F_{\rm obs}(\nu_{\rm obs})= \frac{\delta_D^3(1+z)}{d_L^2} V 
	j_{\rm rad}\left(\frac{1+z}{\delta_D}\nu_{\rm obs},\mu,\phi_{\rm obs}\right)
	\quad {\rm erg/cm}^2{\rm /s/Hz}
\end{align}
where, $z$ is the redshift of the source, $d_L$ is the luminosity distance, $V$ is the volume of the 
emission region, $j_{\rm rad}$ is the emissivity due to synchrotron/SSC/EC process, $\mu$ is the viewing angle
measured from the frame of emission region -- equation (\ref{eq:mu_er}), and $\phi_{\rm obs}$
the azimuthal angle of the observer.  An approximate solution of the observed
flux can be obtained by replacing the emissivity in the above equation with its corresponding
analytical approximation: equation (\ref{eq:syn_approx})/(\ref{eq:ssc_approx})/(\ref{eq:ec_approx}). 
It is then straight forward to obtain the relation between
the source parameters and the observed fluxes due to synchrotron, SSC and EC 
processes as
\begin{align}\label{eq:obs_syn_approx}
	{F}_{\rm obs}^{\rm syn}(\nu_{\rm obs})&\approx\left\{
\begin{array}{ll}
	\mathbb{S}(z,p)\,\delta_D^{\frac{p+5}{2}}B^{\frac{p+1}{2}}
	R^{3}K\nu_{\rm obs}^{-\left(\frac{p-1}{2}\right)}
&\textrm{for}\quad \mbox {~$\nu_{\rm obs}\ll\delta_D\gamma_b^2\nu_L/(1+z)$~} \\
\mathbb{S}(z,q)\,\delta_D^{\frac{q+5}{2}}B^{\frac{q+1}{2}}
R^{3}K\gamma_b^{q-p}\nu_{\rm obs}^{-\left(\frac{q-1}{2}\right)}
	&\textrm{for}\quad \mbox {~$\nu_{\rm obs}\gg\delta_D\gamma_b^2\nu_L/(1+z)$~}
\end{array}
\right. \\
\nonumber \\
	\label{eq:obs_ssc_approx}
	{F}_{\rm obs}^{\rm ssc}(\nu_{\rm obs})&\approx \left\{
\begin{array}{ll}
	\mathbb{C}(z,p)\,\delta_D^{\frac{p+5}{2}}B^{\frac{p+1}{2}}
	R^{4}K^2\nu_{\rm obs}^{-\left(\frac{p-1}{2}\right)}\textrm{log}\left(\frac{\gamma_b}{\gamma_{\rm min}}\right)
&\textrm{for}\quad \mbox {~$\nu_{\rm obs}\ll\delta_D\gamma_b^4\nu_L/(1+z)$~} \\
\mathbb{C}(z,q)\,\delta_D^{\frac{q+5}{2}}B^{\frac{q+1}{2}}
R^{4}K^2\gamma_b^{2(q-p)}\nu_{\rm obs}^{-\left(\frac{q-1}{2}\right)} 
\textrm{log}\left(\frac{\gamma_{\rm max}}{\gamma_b}\right)
	&\textrm{for}\quad \mbox {~$\nu_{\rm obs}\gg\delta_D\gamma_b^4\nu_L/(1+z)$~}
\end{array}
\right. \\
\nonumber \\
\label{eq:obs_ec_approx}
{F}_{\rm obs}^{\rm ec}(\nu_{\rm obs})&\approx \left\{
\begin{array}{ll}
	\mathbb{E}(z,p)\,\delta_D^{p+3}U_*\overline{\nu}_*^{\frac{p-3}{2}}
	R^{3}K\nu_{\rm obs}^{-\left(\frac{p-1}{2}\right)}
	&\textrm{for}\quad \mbox {~$\nu_{\rm obs}\ll\delta_D\Gamma\gamma_b^2\,\overline{\nu}_*/(1+z)$~} \\
	\mathbb{E}(z,q)\,\delta_D^{q+3}U_*\overline{\nu}_*^{\frac{q-3}{2}}
	R^{3}K\gamma_b^{q-p}\nu_{\rm obs}^{-\left(\frac{q-1}{2}\right)}
&\textrm{for}\quad \mbox {~$\nu_{\rm obs}\gg\delta_D\Gamma\gamma_b^2\,\overline{\nu}_*/(1+z)$~}
\end{array}
\right.
\end{align}
Here, $\mathbb{S}$, $\mathbb{C}$ and $\mathbb{E}$ are the quantities involving physical constants, redshift
and particle index. For the EC process, we have assumed cos$\,\theta\sim 1$ and $\beta_{\Gamma}\sim 1$.
In Figure \ref{fig:approx}, we show the observed flux
due to synchrotron, SSC and EC processes (solid lines) for a set of source parameters (described in the caption) along with their
approximate analytical solutions (dashed lines). We find that the approximate analytical solution
of fluxes closely agree with the actual numerical results (except around the peak) and hence can be 
used to estimate the source parameters.

\subsection{Source Parameters}\label{sec:source_par}

It is quite evident from equations  (\ref{eq:obs_syn_approx}), (\ref{eq:obs_ssc_approx}) and 
(\ref{eq:obs_ec_approx}), the observed flux at any frequency basically depends upon $12$ 
source parameters namely, $K$, $\gamma_{\rm min}$, $\gamma_{\rm max}$, $\gamma_b$, $p$, $q$, $\delta_D$, 
$\Gamma$, $B$, $R$, $\overline{\nu}_*$ and $U_*$. Among these, $p$ and $q$ can be easily constrained
from observed spectral indices since, the spectral indices due to synchrotron and 
inverse Compton processes will be $(p-1)/2$ and $(q-1)/2$ (section \S\ref{sec:obs_approx}). 
Now, as the low energy end of the blazar SED is affected by 
synchrotron self absorption and the high energy tail by Klein-Nishina effects (or often 
unknown), it is hard to estimate the parameters $\gamma_{\rm min}$ and $\gamma_{\rm max}$. However,
on the basis of shock acceleration theory, one can impose a constrain on $\gamma_{\rm min}$ such that 
$\gamma_{\rm min}\gtrsim\Gamma$ \citep{2002ApJ...564...97K}. 
On the other hand, $\gamma_{\rm max}$ is a weak parameter and 
can be chosen to reproduce the highest energy of the gamma ray photon observed. 
Thus, after assigning a convenient choice for $\gamma_{\rm min}$ and $\gamma_{\rm max}$, we are 
finally left with $8$ parameters which are to be determined from observations. 

A good spectral information at optical/UV/X-ray energies will let us identify the 
synchrotron peak frequency ($\nu_{\rm sp,obs}$) in the blazar SED and the
same can be expressed in terms of the source parameters as
\begin{align} \label{eq:nus_obs}
	\nu_{\rm sp,obs}=\left(\frac{\delta_D}{1+z}\right)\gamma_b^2 \nu_L
\end{align}
Similarly, if one can identify the SSC and the EC peak from the high energy spectrum, then
the SSC peak can be expressed as 
\begin{align} \label{eq:nuc1_obs}
	\nu_{\rm sscp,obs}=\left(\frac{\delta_D}{1+z}\right)\gamma_b^4 \nu_L
\end{align}
and the EC peak
\begin{align} \label{eq:nuc2_obs}
	\nu_{\rm ecp,obs}=\left(\frac{\delta_D\Gamma}{1+z}\right)\gamma_b^2 \,\overline{\nu}_*
\end{align}
If the external photon field is assumed to be a blackbody, illuminated by the accretion disk, then $U_{\rm ph*}$ and 
$\overline{\nu}_*$ can be related as
\begin{align}\label{eq:ext_freq}
	\overline{\nu}_* = 2.82 \,f_{ext}\, \frac{K_B}{h}\left(\frac{c}{4\sigma_{SB}}\int U_{\rm ph*}(\nu_{i*})\, d\nu_{i*}\right)^{1/4}
\end{align}
where, $K_B$ is the Boltzmann constant, $\sigma_{SB}$ is the Stefan-Boltzmann constant, $ U_{\rm ph*}(\nu_{i*})$ is
the blackbody energy density at frequency $\nu_{i*}$ and $f_{ext}$ is the covering factor describing the fraction of
external photons participating in the inverse Compton process. Besides these, 
we can also express the magnetic field energy density ($U_B$) in terms of electron energy density ($U_e$) as
\begin{align}\label{eq:equipart}
	U_B = \eta U_e
\end{align}
where, 
\begin{align}
	U_B = \frac{B^2}{8\pi}\quad {\rm erg/cm}^3 \quad \textrm{and} \quad U_e = m_e c^2 \int\limits_{\gamma_{\rm min}}^{\gamma_{\rm max}} \gamma N(\gamma)d\gamma \quad {\rm erg/cm}^3 \nonumber
\end{align}
Here, $\eta\approx 1$ corresponds to the equipartition condition indicating total energy of 
the system to be minimum \citep{1970ranp.book.....P}. Hence, the knowledge of $\nu_{\rm sp,obs}$, $\nu_{\rm sscp,obs}$, $\nu_{\rm ecp,obs}$
and the fluxes at Optical (Synchrotron; equation (\ref{eq:obs_syn_approx})), 
X-ray (SSC; equation (\ref{eq:obs_ssc_approx})) and gamma ray (EC; equation (\ref{eq:obs_ec_approx})), 
along with equations (\ref{eq:ext_freq}) and (\ref{eq:equipart}), can in principle, let one
estimate the remaining $8$ source parameters by solving the corresponding coupled equations.

In case of simple models involving only synchrotron and SSC alone (for e.g. SED of many BL Lac objects), 
the total number of source
parameters reduces to $9$ since, the parameters $\Gamma$, $\overline{\nu}_*$ and $U_*$ will be
redundant. Leaving the electron spectral indices, $\gamma_{\rm min}$ and $\gamma_{\rm max}$, we will be 
left with only $5$ parameters which can be estimated from the set of coupled equations 
(\ref{eq:obs_syn_approx}), (\ref{eq:obs_ssc_approx}), (\ref{eq:nus_obs}), (\ref{eq:nuc1_obs})
and (\ref{eq:equipart}). Non-availability of any of these observables may not allow to estimate an unique set of parameters and
one needs to assume certain parameters a priori. Alternatively, one can add other observable 
features (e.g. variability timescale, synchrotron self absorption break frequency, transition frequency from   
dominant synchrotron emission to inverse Compton etc) to constrain the model and obtain a unique set of
source parameters.

\section{XSPEC Spectral fit} \label{sec:xspecfit}

We developed numerical codes to calculate the emissivities corresponding to synchrotron, SSC  and EC
emission processes, which are then used to estimate the observed fluxes after accounting for the 
relativistic and cosmological transformations. 
The codes are optimized by incorporating quadrature integrations and
different interpolation schemes to reduce the run time\footnote{Typically, the runtime
for 1000 iterations of generating 100 flux points sampled logarithmically over a broadband SED, spanning 
over radio to gamma ray energies, and involving synchrotron, SSC and EC emission processes, 
on an Intel i5 machine (3.3 GHz $\times$ 4 processors) with 8 GB RAM, is 4 mins approximately.}.
These codes are then added as additive local models to the XSPEC package following the
standard prescription\footnote{http://heasarc.gsfc.nasa.gov/docs/xanadu/xspec/manual/manual.html}. 
We choose the parameters of the combined XSPEC models as, the electron spectral indices $p$ and $q$, minimum and the 
maximum electron energies $\gamma_{\rm min}$ and $\gamma_{\rm max}$, synchrotron
peak frequency $\nu_{\rm sp,obs}$, SSC peak frequency $\nu_{\rm sscp,obs}$, EC peak frequency $\nu_{\rm ecp,obs}$,
synchrotron flux ${F}^{\rm syn}_{\rm obs}$ at a reference frequency $\nu_{\rm syn,obs}^{\rm ref}$, SSC flux ${F}^{\rm ssc}_{\rm obs}$ at a reference frequency 
$\nu_{\rm ssc,obs}^{\rm ref}$, EC flux ${F}^{\rm ec}_{\rm ec,obs}$ at a reference frequency $\nu_{\rm ec,obs}^{\rm ref}$ and 
equipartition factor $\eta$.
To extend the application of the code to fit the SED of misaligned AGNs, we also include an
option to incorporate large viewing angles. This is achieved by considering the ratio of the Doppler factor
$\delta_D$ to the bulk Lorentz factor $\Gamma$ as an additional parameter.
The advantage of providing observed parameters as input to the XSPEC codes will let us to avoid the 
uncertainty regarding the correct choice of initial guess parameters as well as
facilitate a faster convergence.
These observed parameters are then converted into source parameters within the code 
by solving the approximate coupled equations and other conditions described in the earlier section. 
Consistently, the same procedure is then used to extract the best fit source parameters from the fitted observational quantities.

\subsection{Spectral fitting of 3C\,279}

\begin{figure}
\centerline{\includegraphics[angle=-90,width=12cm]{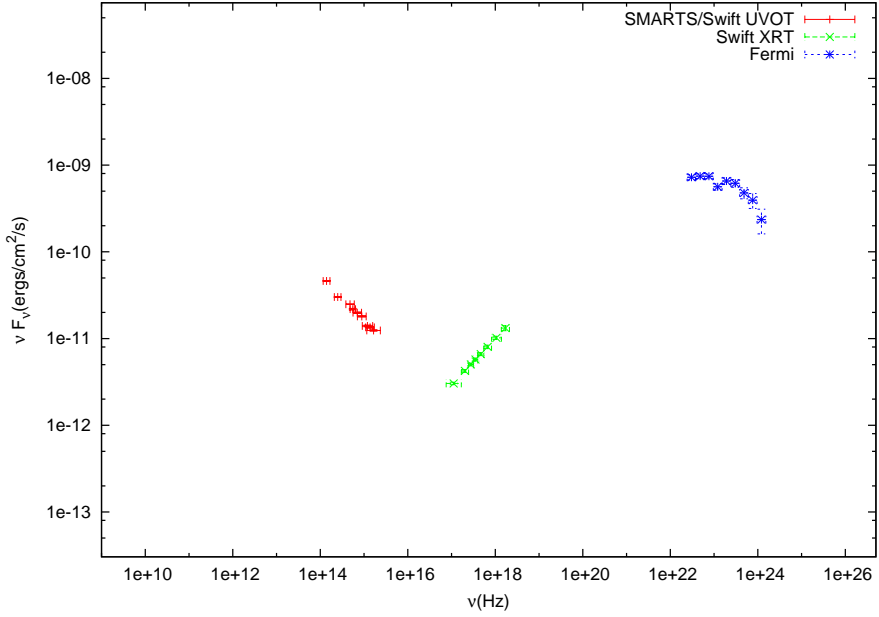}}
\caption{Broadband SED of 3C\,279 during its gamma ray high state on 2-8 April 2014. The 
source was simultaneously observed at optical/UV (\emph{SMARTS}, \emph{Swift-UVOT}), X-ray (\emph{Swift-XRT})
and gamma ray (\emph{Fermi}) energies  \citep{2015ApJ...803...15P}.
\label{fig:3c279obs}}
\end{figure}

To further study and validate the proposed broadband spectral fitting algorithm using 
XSPEC, we choose the well studied FSRQ, 3C\,279 ($z=0.536$), as a test case. We select the flaring 
epoch of 3C\,279, during March-April 2014, when the source was observed to be very bright in 
gamma rays. This huge gamma ray flare was witnessed by \emph{Fermi} gamma ray telescope and was
simultaneously monitored at X-ray energies by \emph{Swift-XRT} and in UV/optical by \emph{Swift-UVOT},
\emph{SMARTS} and \emph{Steward} observatories, thereby providing an unprecedented multi wavelength
data \citep{2015ApJ...803...15P}. In Figure \ref{fig:3c279obs}, we show the observed SED corresponding to the highest gamma ray flux state 
(2-8 April 2014) encountered during this flaring episode.

Earlier studies on the broadband SED of 3C\,279 suggests, substantial contribution of synchrotron, 
SSC and EC processes and this further assures that this source can be a right choice for 
testing the proposed spectral fitting algorithm \citep{2012MNRAS.419.1660S,2001ApJ...553..683H}. The justification behind this inference is that 
the observed X-ray and gamma ray fluxes  from 3C\,279 cannot be interpreted under single emission process
like SSC or EC, as it demands a magnetic field that deviate largely from the equipartition
condition. In addition, detection of 3C\,279 at very high energy gamma rays (VHE)
with relatively hard spectrum
indicates, the EC process to be dominated by scattering of infrared photons from the obscuring
torus (EC/IR), rather than the Lyman alpha line emission from the broad line emitting 
regions (EC/BLR) \citep{2009MNRAS.397..985G}. For the flaring period under consideration, no significant
detection of VHE emission was reported from the source and hence, we cannot assert that 
the gamma ray emission to be an outcome of EC/IR or EC/BLR processes. However, it can be shown that the observed
fluxes at optical, X-ray and gamma ray energies support EC/IR interpretation of the high energy
emission \citep{2017arXiv170506185S}.

\begin{figure}
\centerline{\includegraphics[angle=-90,width=15cm]{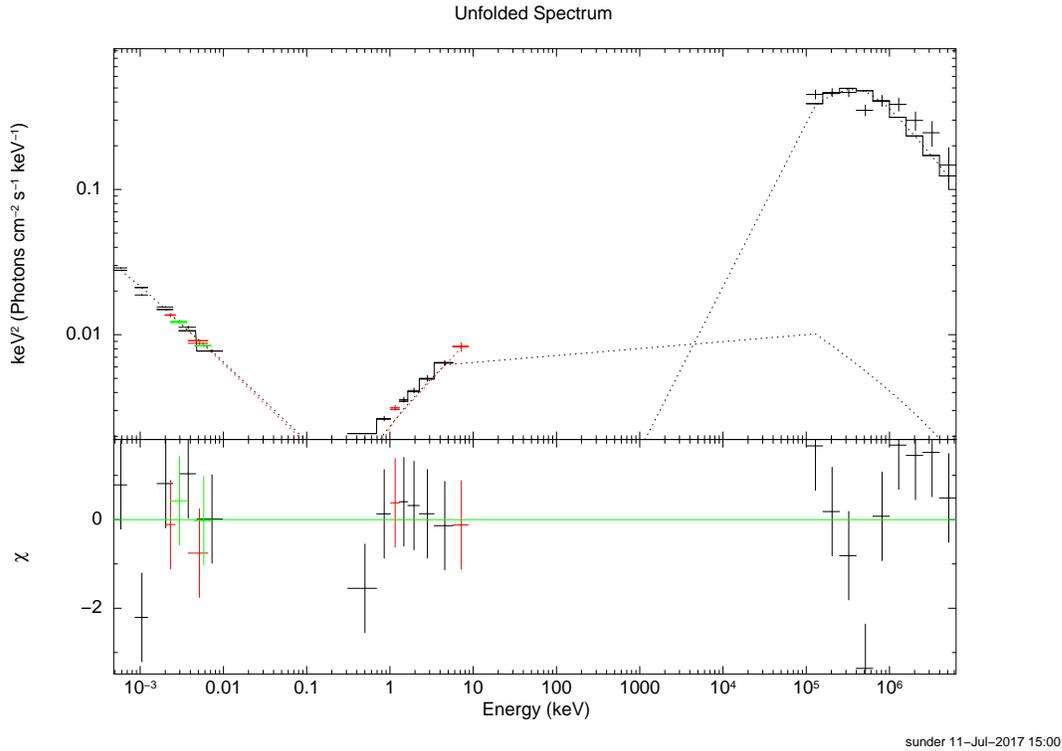}}
\caption{XSPEC spectral fit of the broadband SED of 3C\,279 using synchrotron, SSC and EC processes.
\label{fig:3c279fit}}
\end{figure}

\begin{figure}
\centering
\begin{tabular}{@{}cc@{}}
	{\includegraphics[angle=-90,width=0.5\textwidth]{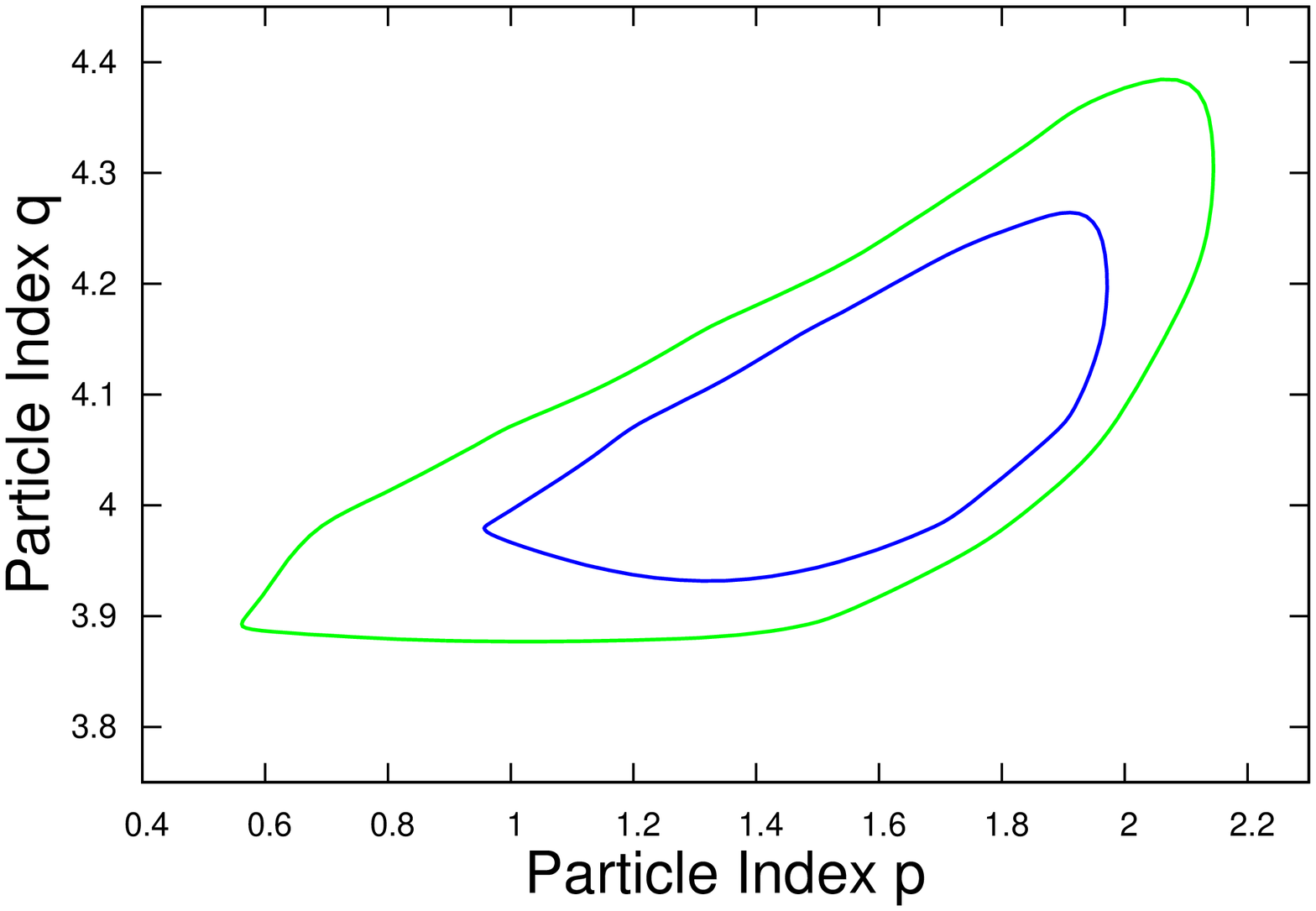}}&
	{\includegraphics[angle=-90,width=0.5\textwidth]{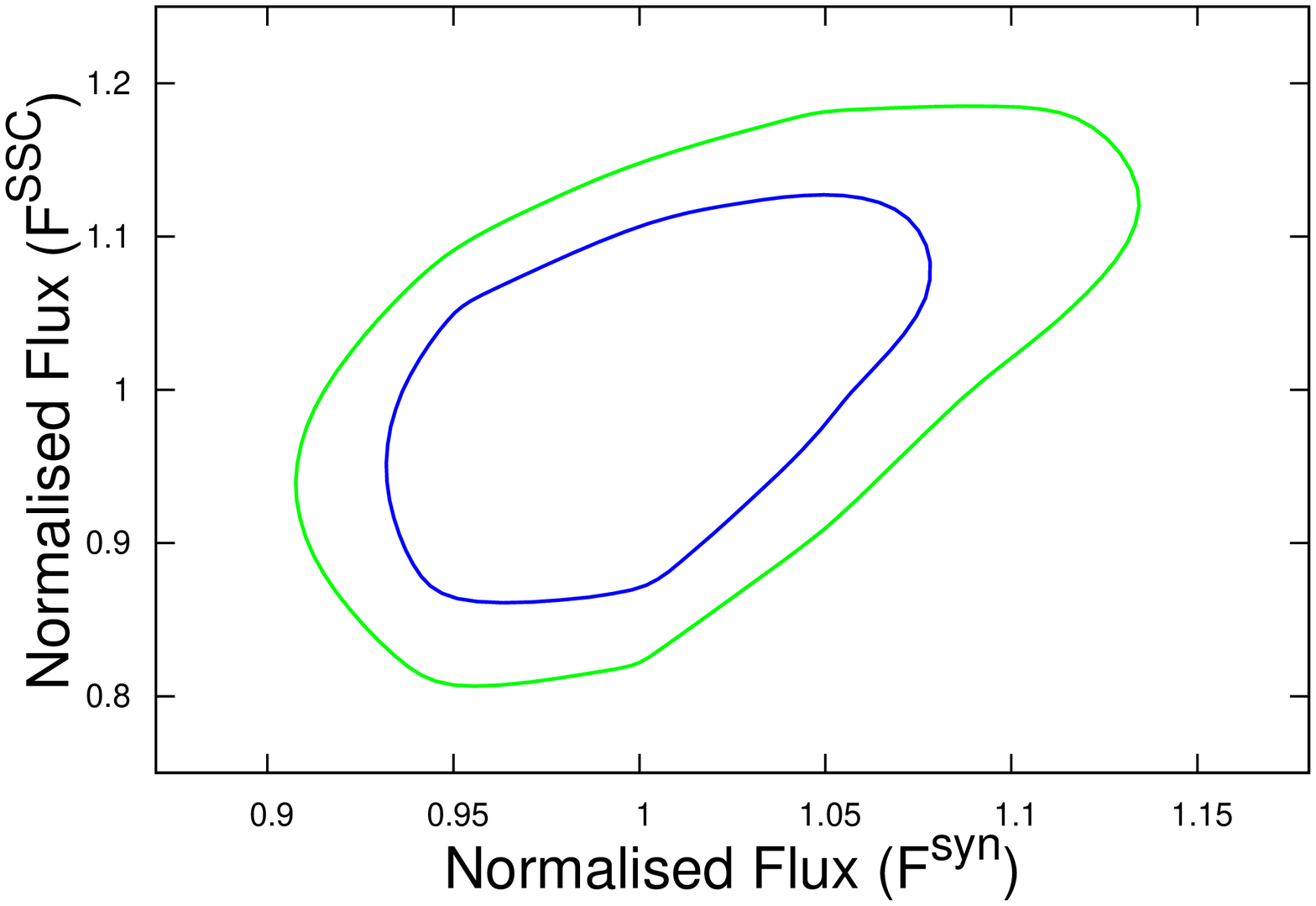}}
\end{tabular}
\caption{The 1-$\sigma$ (blue) and 2-$\sigma$ (green) confindence interval between the broken power law 
electron spectral indices $p$ and $q$ (left), and the synchrotron and SSC fluxes normalised to its
best fit flux (right).
\label{fig:obscont1245}}
\end{figure}

\begin{figure}
\centerline{\includegraphics[angle=-90,width=0.5\textwidth]{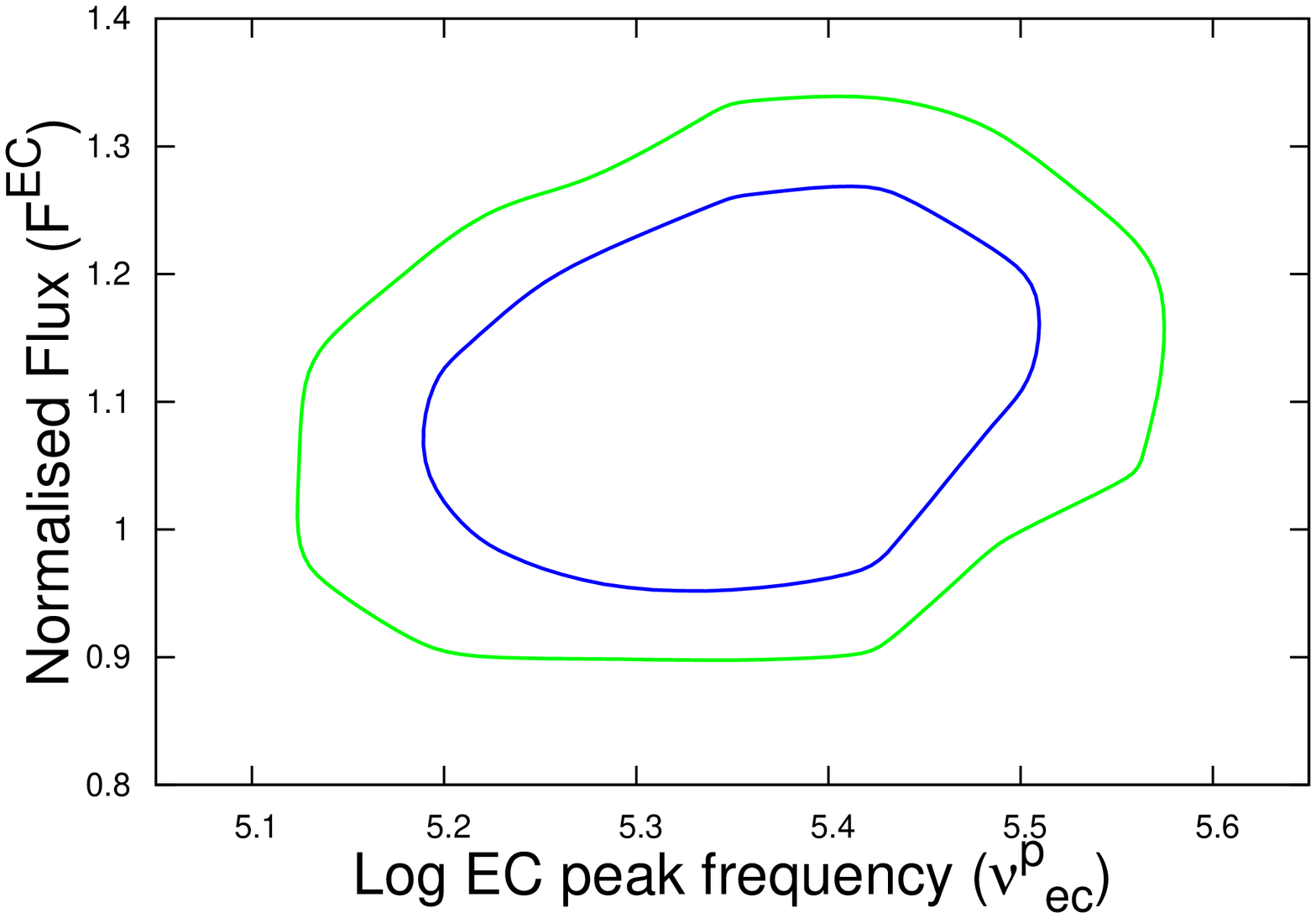}}
\caption{The 1-$\sigma$ (blue) and 2-$\sigma$ (green) confindence interval between the EC peak 
frequency and the normalised EC flux.
\label{fig:obscont36}}
\end{figure}

Though the flare under consideration was simultaneously monitored at optical, X-ray and gamma ray 
energies, non-availability of lower frequency observation at microwave/IR prevents us from 
estimating $\nu_{\rm sp,obs}$. Similarly, $\nu_{\rm sscp,obs}$ also remains uncertain since the the X-ray 
spectra do not show any signature of a peak. A lack of these informations causes a deficit 
in the number of observables and thereby, prevents us from obtaining a unique set of source parameters. 
Thus, we fix the values of $\nu_{\rm sp,obs}$ and $\nu_{\rm sscp,obs}$ to appropriate values to obtain meaningful source parameters. Accordingly, these quantities are fixed at $\nu_{\rm sp,obs}=3.8\times 10^{13}$ Hz 
and $\nu_{\rm sscp,obs}=7.6 \times 10^{19}$ Hz, and we fitted the spectrum to obtain the rest of the 
observables. Further, to allow for the uncertainties regarding the 
emission models, a systematic error of 10\% was applied evenly on all the emission models in addition to 
the uncertainties in the observed fluxes. Finally, the best fit spectrum along with the residual, 
resulting from the present study, is shown in Figure \ref{fig:3c279fit}. 
In Table \ref{tab:obspar}, we give the best fit observational
quantities corresponding to a minimum reduced chi square of $\chi_{red}=0.8$ for $20$ degrees of freedom. 
The $1$-sigma confidence range of these
quantities are obtained by scanning the parameter space around this minima. In Figures 
\ref{fig:obscont1245} and \ref{fig:obscont36}, we show the 
contour plots between different quantities for 1-$\sigma$ ($\Delta \chi^2=2.3$) 
and 2-$\sigma$ ($\Delta \chi^2=4.61$) confidence levels. 

\begin{table}
\bc
\begin{minipage}[]{100mm}
	\caption[]{XSPEC fit result\label{tab:obspar}}\end{minipage}
\setlength{\tabcolsep}{1pt}
\small
\begin{tabular}{lcc}
	\hline\noalign{\smallskip}
      Observable & {Symbol} & {Value} \\
				 \hline\noalign{\smallskip}
        Low energy Particle index    &  $p$      &    1.64   \\\\
        High energy Particle index    &     $q$    &    4.09   \\\\
		Synchrotron peak frequency$^f$ (Hz)    &     $\nu^p_{\rm syn}$  &   $3.83\times10^{13}$   \\\\
		SSC peak frequency$^f$ (Hz)    &     $\nu^p_{\rm ssc}$  &   $7.65\times10^{19}$  \\\\
		EC peak frequency (Hz)    &     $\nu^p_{\rm ec}$  &   $5.0\times10^{22}$   \\\\
		Synchrotron Flux  (erg/cm$^2$/s)  &     $F^{\rm syn}$   &   $3.23\times10^{-11}$   \\\\
		Synchrotron reference frequency$^*$ (Hz)& $\nu_{\rm syn}^{ref}$& $2.4\times10^{14}$\\\\
		SSC Flux (erg/cm$^2$/s)  & $F^{\rm syn}$ & $9.88\times10^{-12}$\\\\
		SSC reference frequency$^*$ (Hz)&$\nu_{\rm ssc}^{ref}$ & $4.79\times10^{17}$\\\\
		EC Flux (erg/cm$^2$/s)  & $F^{\rm syn}$ & $3.44\times10^{-10}$\\\\
		EC reference frequency$^*$ (Hz)&$\nu_{\rm ec}^{ref}$ & $4.79\times10^{23}$\\\\
		Equipartition factor$^f$& $\eta$ &0.1\\\\
		Ratio of Doppler to Lorentz factor$^f$& $\delta_D/\Gamma$& 1\\\\
	    Minimum electron energy$^f$ & $\gamma_{\rm min}$      &   40   \\\\
	    Maximum electron energy$^f$ & $\gamma_{\rm max}$      &   $10^6$ \\
		\noalign{\smallskip}\hline
    \end{tabular}
	\ec
      \tablecomments{0.86\textwidth}{Best fit observable quantities/source parameters of 3C~279, during the gamma ray flare on 2014, 
		  obtained using XSPEC emission models developed in this work. Quantities with superscript $f$ are fixed and 
	  not included in the fitting. The reference frequencies, denoted by superscript $*$, are the ones at which the 
observed fluxes are fitted.} 
\end{table}

\begin{figure}
\centering
\begin{tabular}{@{}cc@{}}
	{\includegraphics[angle=-90,width=0.5\textwidth]{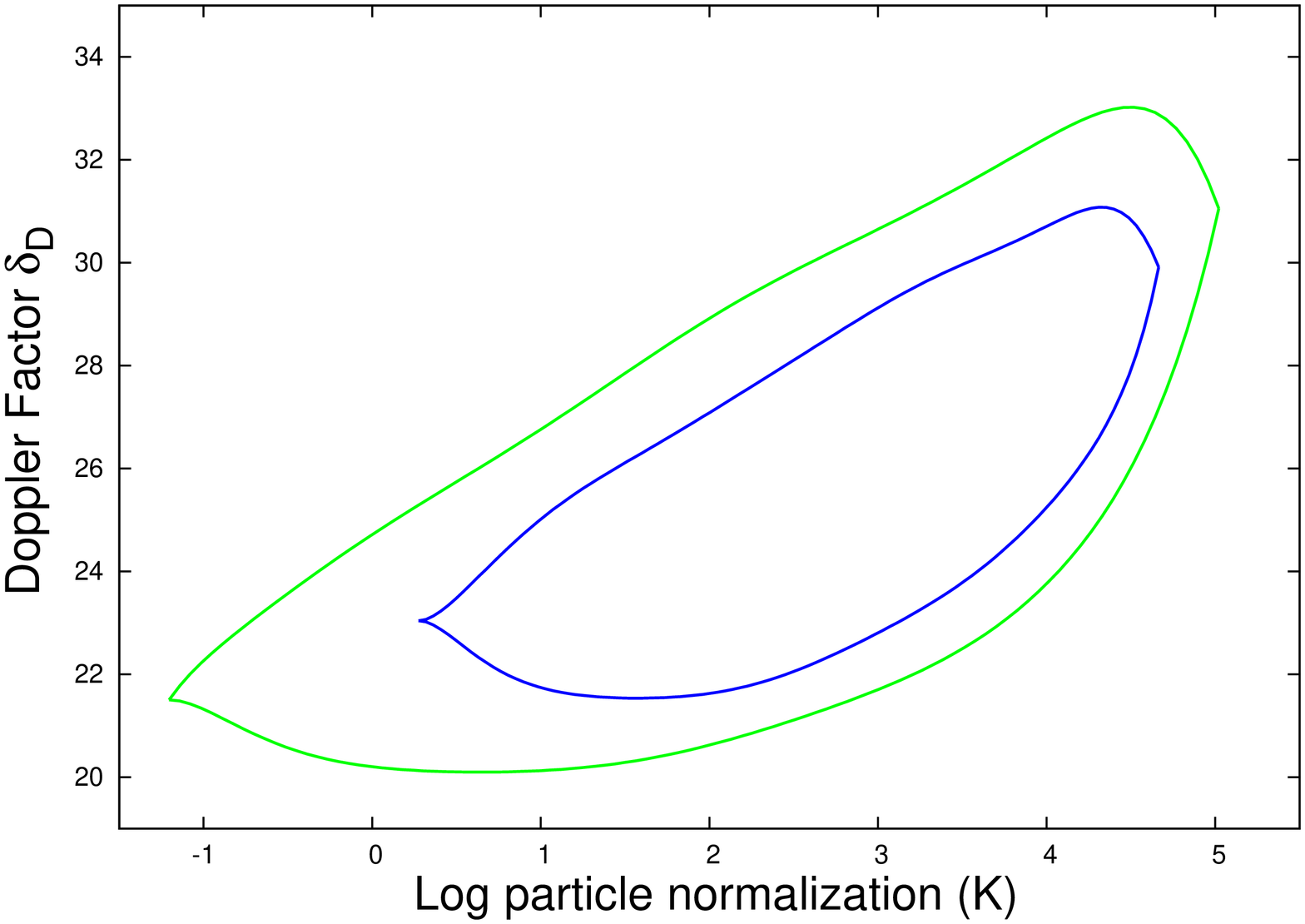}}&
	{\includegraphics[angle=-90,width=0.5\textwidth]{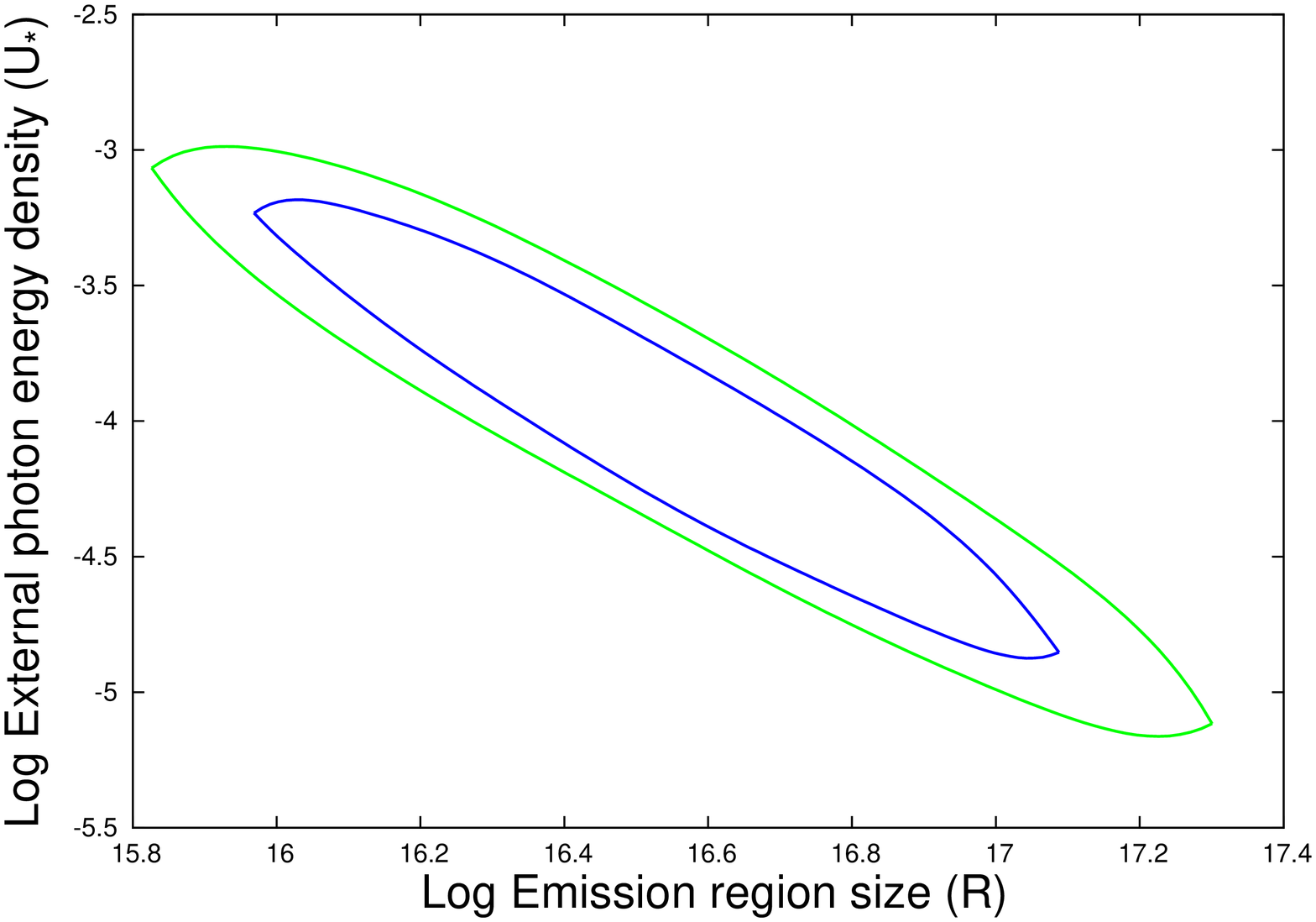}} \\
\end{tabular}
\caption{The 1-$\sigma$ (blue) and 2-$\sigma$ (green) confindence interval  
between the particle normalisation $K$ (in log) and the Doppler factor $\delta$ is shown 
at the left; whereas, the
one at right is between emission region size $R$ (in log) and the external photon energy 
density $U_*$ (in log). 
\label{fig:physcont}}
\end{figure}

\begin{figure}
\centerline{\includegraphics[angle=-90,width=0.5\textwidth]{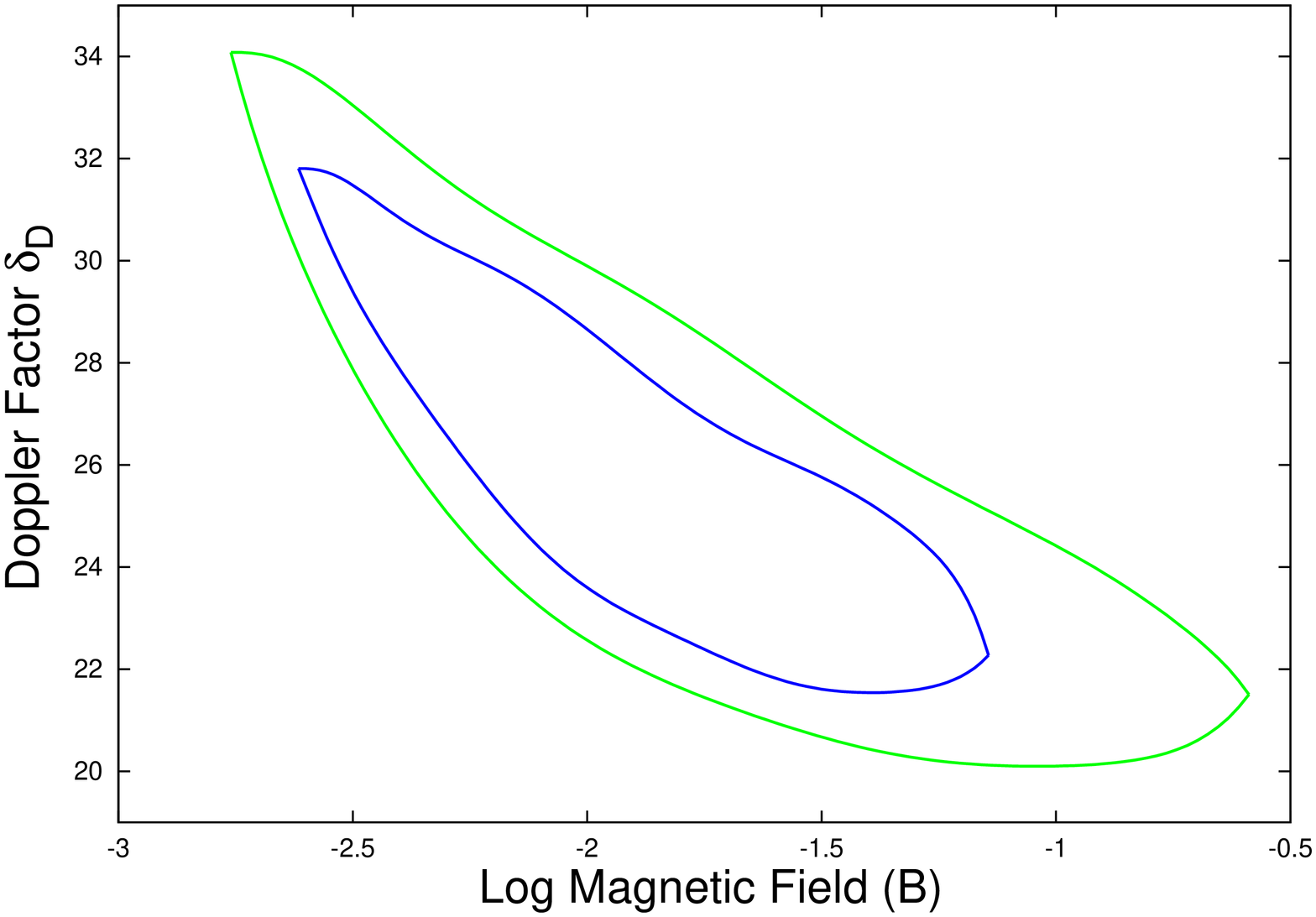}}
\caption{The 1-$\sigma$ (blue) and 2-$\sigma$ (green) confindence interval between the magnetic field $B$
(in log) and the Doppler factor $\delta$.
\label{fig:physcontbdel}}
\end{figure}

The knowledge of the best fit observational quantities can be inverted back to obtain  
the corresponding source parameters using the approximate analytical expression described 
earlier (\S\ref{sec:obs_approx} and \S\ref{sec:source_par}). 
Since the emission codes use the same expressions to derive the 
source parameters and the emissivities, the resulting source parameters will also be the
best fit values giving rise to same $\chi^2$. In Table \ref{tab:physpar}, we give the
source parameters derived from the best fit observable quantities, mentioned in Table \ref{tab:obspar}.
To obtain the confidence range, we again use the approximate analytical 
expressions to extract the source parameter range from the observable parameter space. However to be consistent with the 
freezing of the observed quantities $\nu_{\rm sp,obs}$ and  $\nu_{\rm sscp,obs}$, we fix 
the source parameters $\overline{\nu}_* = 6\times 10^{13}$ Hz  (corresponding to $T_*\approx 1000$ K) and $\gamma_b=1.4 \times 10^3$. 
In Figure \ref{fig:physcont} and \ref{fig:physcontbdel}, we show the 
contour plots between the rest of the source parameters namely, $\delta_D$, $K$, $U_*$, $B$ and $R$, 
corresponding to 1-$\sigma$ and 2-$\sigma$ confidence levels.

\begin{table}
	\bc
\begin{minipage}[]{100mm}
	\caption[]{Best fit source parameters\label{tab:physpar}}\end{minipage}
\setlength{\tabcolsep}{1pt}
\small
    \begin{tabular}{lcc}
	\hline\noalign{\smallskip}
      Observable$^\mathrm{a}$ & \multicolumn{1}{c}{Symbol}
                 & \multicolumn{1}{c}{Value} \\\hline
        Low energy Particle index    &  $p$      &    1.64   \\\\
        High energy Particle index    &     $q$    &    4.09   \\\\
		Particle normalisation  (cm$^{-3}$)  &     $K$  &   $2.45\times10^{3}$   \\\\
		Break Lorentz factor    &     $\gamma_b$  &   $1.41\times10^{3}$  \\\\
		Minimum electron Lorentz factor    &     $\gamma_{\rm min}$  &   $40$   \\\\
		Maximum electron Lorentz factor    &     $\gamma_{\rm max}$  &   $1.0\times10^{6}$   \\\\
		Bulk Lorentz factor& $\Gamma$ & $25.45$\\\\
		Doppler Factor  &     $\delta_D$   &   $25.45$   \\\\
		Magnetic Field (G)  & $B$ & $0.41$\\\\
		Emission region size (cm)&$R$ & $2.36\times10^{16}$\\\\
		Target photon frequency (Hz)  & $\overline{\nu}_*$ & $5.95\times10^{13}$\\\\
		Target photon energy density (erg/cm$^3$)&$U_*$ & $1.88\times10^{-4}$\\
		\noalign{\smallskip}\hline
    \end{tabular}
	\ec
      \tablecomments{0.86\textwidth}{The source parameters corresponding to the best fit observable 
		  quantities given in Table \ref{tab:obspar}. These values are extracted using the same approximate 
	  analytical expressions used in the XSPEC emission models.} 
\end{table}


\section{Discussion} \label{sec:discussion}

The blazar spectral fitting algorithm demonstrated in the present work provides a convenient
way to understand the different emission processes as well as to extract the parameters
governing the source. The error ellipses between different parameters 
(Figures \ref{fig:obscont1245}, \ref{fig:obscont36}, \ref{fig:physcont} and \ref{fig:physcontbdel}), indicate 
the allowed ranges and the possible correlations between the parameters. Availability of a well sampled
SED of a source at synchrotron, SSC and EC spectral components, will let one to perform the fitting 
with more free parameters. This in turn, will help us to understand the physical condition of the source
during that particular observation.

Besides providing the best fit parameters, the algorithm developed in this work, will also help us 
to eliminate the degenerate parameters. Lack of information about the 
observed quantities, like peak frequencies, fluxes due to different emission processes, etc., will 
lead to degenerate source parameters irrespective of having a well sampled data. In 
conventional algorithms, where fitting is performed directly on the source parameters, this 
degeneracy between the parameters cannot be anticipated and can lead to misconceptions.
For example, a spectral fit similar to the one shown in Figure \ref{fig:3c279fit} 
can be obtained for a different choice of $\nu_{\rm sp,obs}$. However, this will gives rise 
to a different set of source parameters and particularly the target photon 
temperature. In such cases, one cannot differentiate between the target photon field 
responsible for the gamma ray emission through EC scattering. The knowledge of the 
synchrotron peak frequency can, thereby, help us in removing this degeneracy. 
Alternatively, detection of the source at VHE, can also impose certain constraints on the 
temperature of the external photon field \citep{2009MNRAS.397..985G}. Nevertheless, 
the constraints as well as the degeneracy of the parameters depend on the choice of the physical model, 
the initial assumptions and the quality of the observed SED.

The procedure of extracting the physical parameters of the 3C\,279 using approximate analytical 
expressions, without statistical fitting, was also demonstrated by \cite{2012MNRAS.419.1660S}, during 
the flare observed on 2006. They show the high energy emission can be successfully explained by the 
EC scattering of the IR photons and the parameters quoted are comparable to the one presented here.
The observed SED used in the present work was taken from \cite{2015ApJ...803...15P} where, the broadband
SED of the same epoch was modelled using synchrotron, SSC, EC/IR and EC/BLR emission processes. The
quoted parameters differ from the ones obtained here since, the inclusion
of additional emission process will increase the number of parameters which cannot be effectively 
constrained using the limited information available. Nevertheless, the SEDs during the flaring state and
quiescent state can be reproduced satisfactorily under these emission models. \cite{2016MNRAS.456.2173Y}
employed Markov chain Monte Carlo technique to build 14 bright SEDs of 3C\,279. Their emission model
is similar to the one used by \cite{2015ApJ...803...15P}; however, they are able to provide the confidence 
ranges of the obtained parameters corresponding to the adapted Bayesian statistics. \cite{2016MNRAS.457.3535Z}
used an inhomogeneous jet model \citep{2012MNRAS.423..756P} to model the SED of 3C\,279. The jet is assumed
to be conical and the source parameters are chosen to vary along the jet.  Using this model they were able to 
reproduce the broadband SEDs of the source during 2008 and 2010.

Having developed an algorithm to perform a spectral fitting using synchrotron, SSC and EC processes,
the present work can be easily extended to include more than one EC processes \citep{2014ApJ...782...82D}
or reduced to a simple model involving only synchrotron and SSC processes. For the latter case,
the reduction in the number of source parameters (\S\ref{sec:source_par}) and omission of the EC component
of the code will eventually led to faster convergence of the fitting process. Besides the observational quantities 
used in this work for fitting the data, the 
variability time scale can also play an important role in constraining the parameters. Knowledge of the variability
time scale, $t_{\rm var}$, can effectively constrain the size of the emission region as
\begin{align}\label{eq:t_var}
	R \lesssim \frac{c\delta_D\Delta t_{\rm obs}}{c}
\end{align}
Inclusion of this will allow us to omit the equipartition 
condition (equation (\ref{eq:equipart})) for parameter estimation. On the other hand, the obtained parameters can also be used
to verify this condition or to constrain $\gamma_{\rm min}$. 

The treatment described in this work can be modified/improved further by including other observational features
of blazar SEDs. For example, one can include the synchrotron self absorption frequency which can 
effectively constrain the magnetic field. Similarly, transition frequency where the synchrotron emissivity is equal
to the inverse Compton emissivity can be an additional information. This along with other equations can be 
useful in identifying the electron energies responsible for the emission at a given frequency. This is expected to play an 
important role in understanding the evolution of the light curves at different frequencies, the temporal evolution
of the particle distribution and the dynamics of the AGN jets.

\section{Conclusion}
In the present work, we develop a statistical fitting procedure of the broadband 
spectrum of blazars, considering synchrotron, SSC and EC emission mechanisms. To avoid the difficulty
of choosing the initial guess values 
as well as to warrant a faster convergence, we fit the observed quantities, like the peak frequencies,
fluxes due to different emission processes etc, instead of the source parameters governing the observed
spectrum. The source parameters are then calculated using approximate analytical solutions of 
the various emissivities. Finally, we test and
validate the procedure by fitting the simultaneous broadband observation of the FSRQ, 3C\,279, during 
its gamma ray high state. We show that the proposed spectral fitting procedure is successful in extracting 
most of the parameters of the source. In addition, the proposed methodology will be particularly important
for the ongoing/upcoming multiwavelength campaigns which can effectively probe blazars at various energies and 
provide substantial information necessary to extract the probable physical scenario of the source.

\bibliographystyle{raa}
\bibliography{ref}

\end{document}